\documentclass{PoS}
\usepackage{amsmath}
\usepackage{graphicx}
\usepackage{subcaption}
\usepackage{wrapfig}
\usepackage{url}
\title{Towards Ultimate Parton Distributions from p+p and e+p Collisions}

\ShortTitle{Towards Ultimate Parton Distributions from p+p and e+p Collisions}

\author{\speaker{Shaun Bailey}\\
        Rudolf Peierls Centre for Theoretical Physics, University of Oxford, Clarendon Laboratory, Parks Road, Oxford OX1 3PU, United Kingdom\\
        E-mail: \email{shaun.bailey@physics.ox.ac.uk}}

\abstract{We present results from a detailed assessment of the ultimate constraining power of LHC data on the PDFs that can be expected from the complete dataset, in particular after the High-Luminosity (HL) phase. To achieve this, HL-LHC pseudo-data for different projections of the experimental uncertainties are generated, and the resulting constraints on the PDF4LHC15 set are quantified by means of the Hessian profiling method. We find that HL-LHC measurements can reduce PDF uncertainties by up to a factor of 2 to 4 in comparison to state-of-the-art fits, leading to few-percent uncertainties for important observables such as the Higgs boson transverse momentum distribution via gluon-fusion. Our results illustrate the significant improvement in the precision of PDF fits achievable from hadron collider data alone. In addition, we apply the same methodology to the final anticipated data sample from the proposed LHeC, and compare these with the HL-LHC projections, demonstrating an encouraging complementarity between the projected HL-LHC and LHeC constraints.}

\FullConference{XXVII International Workshop on Deep-Inelastic Scattering and Related Subjects - DIS2019\\
		8-12 April, 2019\\
		Torino, Italy}

\begin{document}
\section{Introduction}
\noindent Starting around 2025, a new high-luminosity phase of the LHC is scheduled to begin, bringing a ten-fold increase to the integrated luminosity. With this comes improved statistics which is predicted to have significant impact on a number of SM and BSM studies. In this regard, analyses have been preformed to assess the state of high-energy physics after this phase has been completed, many of which are presented in the HL/HE-LHC Yellow Report \cite{yellow_report}. Included in these is our study of Parton Distribution Functions (PDFs) at the HL-LHC, which has a two-fold impact: first, we are able to estimate the HL-LHC's impact on our understanding of the underlying proton structure, further, we propagate this error into other SM and BSM analysis of physics at the HL-LHC. We provide an overview of this study with a brief extension to the proposed LHeC A more detailed analysis can be found in \cite{hllhc}.
\section{Data Sets and Error Generation}
\begin{wrapfigure}{R}[-0\textwidth]{0.5\textwidth}
    	        \includegraphics[width=0.5\textwidth]{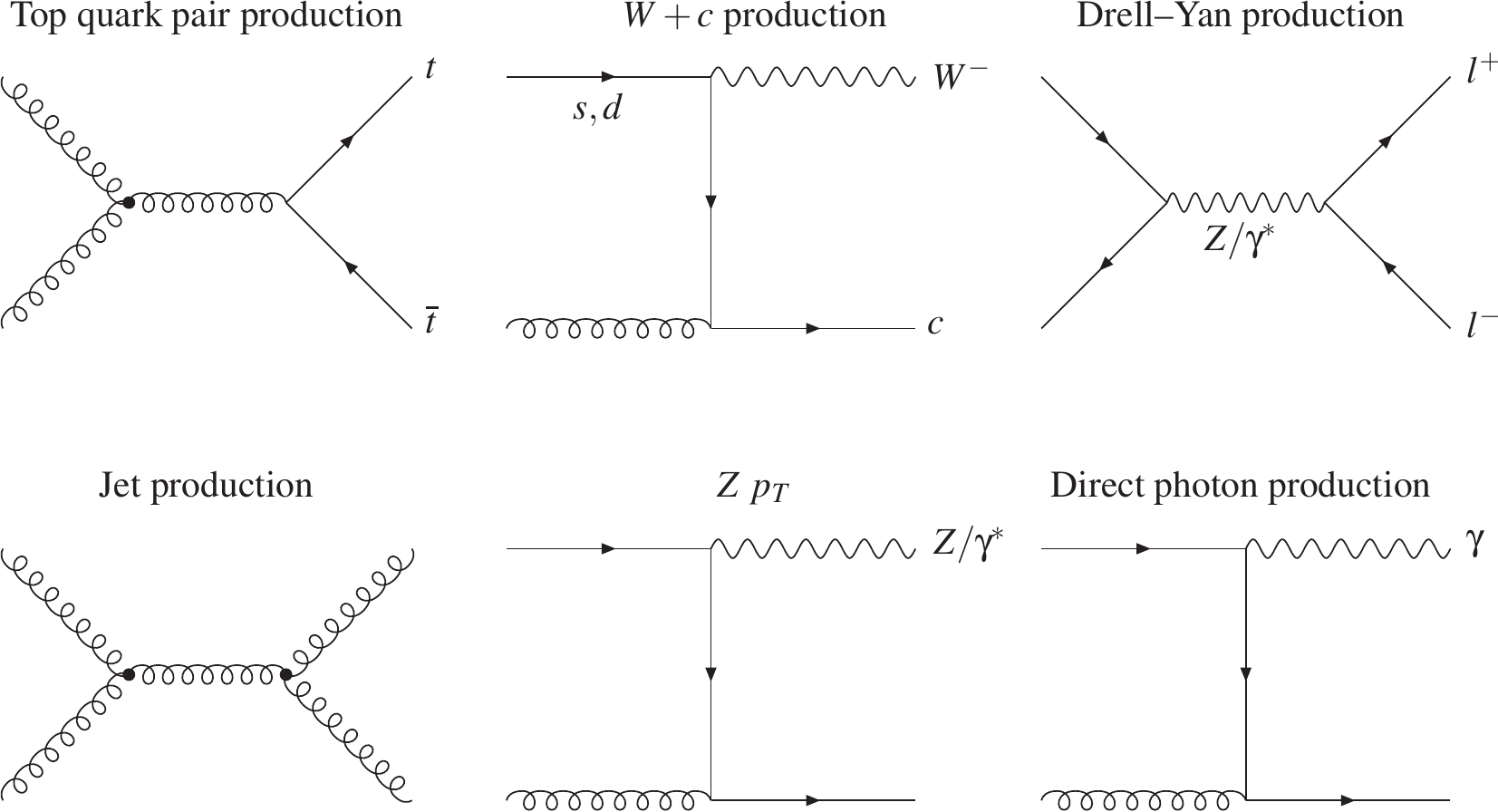}
    	    \caption{LO diagrams for the processes included in the HL-LHC pseudo-data. }
    	    \label{feynman}
\end{wrapfigure}

\noindent We begin by choosing the data sets to use in our analysis, which should be geared towards the HL-LHC strengths, namely the increased statistics. Thus they should roughly fall into two categories: those for which the process is rare and those which will benefit from an extended kinematic reach. We further concentrate on a selection of processes that constrain a variety of PDFs and $x$ values, in particular focusing on the mid-high $x$ region. To this end we choose the processes displayed in figure \ref{feynman}.

\noindent Next we produce corresponding pseudo-data by first generating the theoretical predictions using MCFM \cite{MCFM} interfaced with applgrid \cite{applgrid} (with the exception of jet data which is generated using NLOJET++ \cite{NLOJET}), then predict the final cross section with the PDF4LHC15\_100 PDF set \cite{pdf4lhc}. We use NLO theory for this, as higher order effects are not likely to impact this closure test. For each data point, we then simply shift the theory value randomly according to the predicted error.

\noindent For the statistical errors, we simply take the error to be proportional to $\sqrt{N_\text{obs}}$ where $N_\text{obs}$ is the predicted number of observed events. We  include an acceptance correction that takes into account the effects of branching ratios and detector efficiencies. This is in general extracted from an existing data set for the same process.

\noindent For the systematic errors, we again take these from a previous data set, however we remove the correlations as a full model for this is beyond the scope of this study. Further, as we are concentrating on processes expected to benefit most from improved statistics, we do not expect the detailed modelling of the systematics to have a large impact on the final result. However, by decorrelating the errors, we are artificially reducing their effect. To combat this, we introduce a data-driven correction factor $f_\text{corr} = 0.5$ by which we reduce the uncorrelated systematic errors. We also introduce an additional factor $f_\text{red}$, which parametrises how much we expect the systematic errors to be reduced at the HL-LHC. This improvement could come from upgrades to the detector, better analysing tools, or simply due the increased statistics allowing for better calibration. There are a couple of exceptions, most notably the luminosity is still treated as a correlated 1.5\% error between each point associated with the same detector and uncorrelated with all other detectors.

\section{Hessian Profiling}

\noindent We will be using the Hessian Profiling method for analysing the data, using the results from \cite{hessian}. This is an approximate method that estimates the effect of adding in new data to a PDF set described by Hessian errors. This method assumes that the PDFs stay close to the original when profiling, and thus will be accurate in the context of a closure test.

\noindent The figure of merit is generated by expanding the theoretical predictions:
\begin{align}
\chi^2\!=\!\sum_{i,j=1}^{N_\text{dat}} \left( \sigma^\text{exp}_i - \sigma_{i,0}^\text{th} - \sum_{\alpha = 1}^{N_\text{th}} \Gamma_{i\alpha}^{\text{th}} \beta_{\alpha}^\text{th} \right) \left( \text{cov} \right)_{ij}^{-1} \left( \sigma^\text{exp}_j - \sigma_{j,0}^\text{th} - \sum_{\gamma = 1}^{N_\text{th}} \Gamma_{j\gamma}^{\text{th}} \beta_{\gamma}^\text{th} \right)  + T^2\sum_{\gamma = 1}^{N_\text{th}} \left( \beta_{\gamma}^\text{th} \right)^2,
\label{eq}
\end{align}
where $\sigma^\text{exp}_i$ are the pseudo-data points, $\sigma^\text{th}_{i,0}$ are the theory predictions evaluated using the central PDF, $\beta_{\alpha}^\text{th}$ define the parameter space, $\Gamma_{i,\alpha}^{\text{th}}$ describe how the theory predictions change within this parameter space, $\left( \text{cov} \right)_{ij}$ is the experimental covariance matrix, $N_\text{dat}$ is the number of pseudo-data points and $N_\text{th}$ is the number of theory parameters. The tolerance, $T$, parametrises the effect of the data sets already included in the initial PDF set and is not necessary unity due to tensions and inconsistencies between and within data sets. We take a value of 3, guided by MMHT and CT. We are then able to minimise this to generate our new central PDF. The errors are then calculated using the Hessian corresponding to the above $\chi^2$ in the usual way.

\section{HL-LHC Results}

\subsection{Individual Data Sets}

\begin{figure}[t]
\centering
\begin{subfigure}[b]{0.4\textwidth}
 \includegraphics[width=\textwidth]{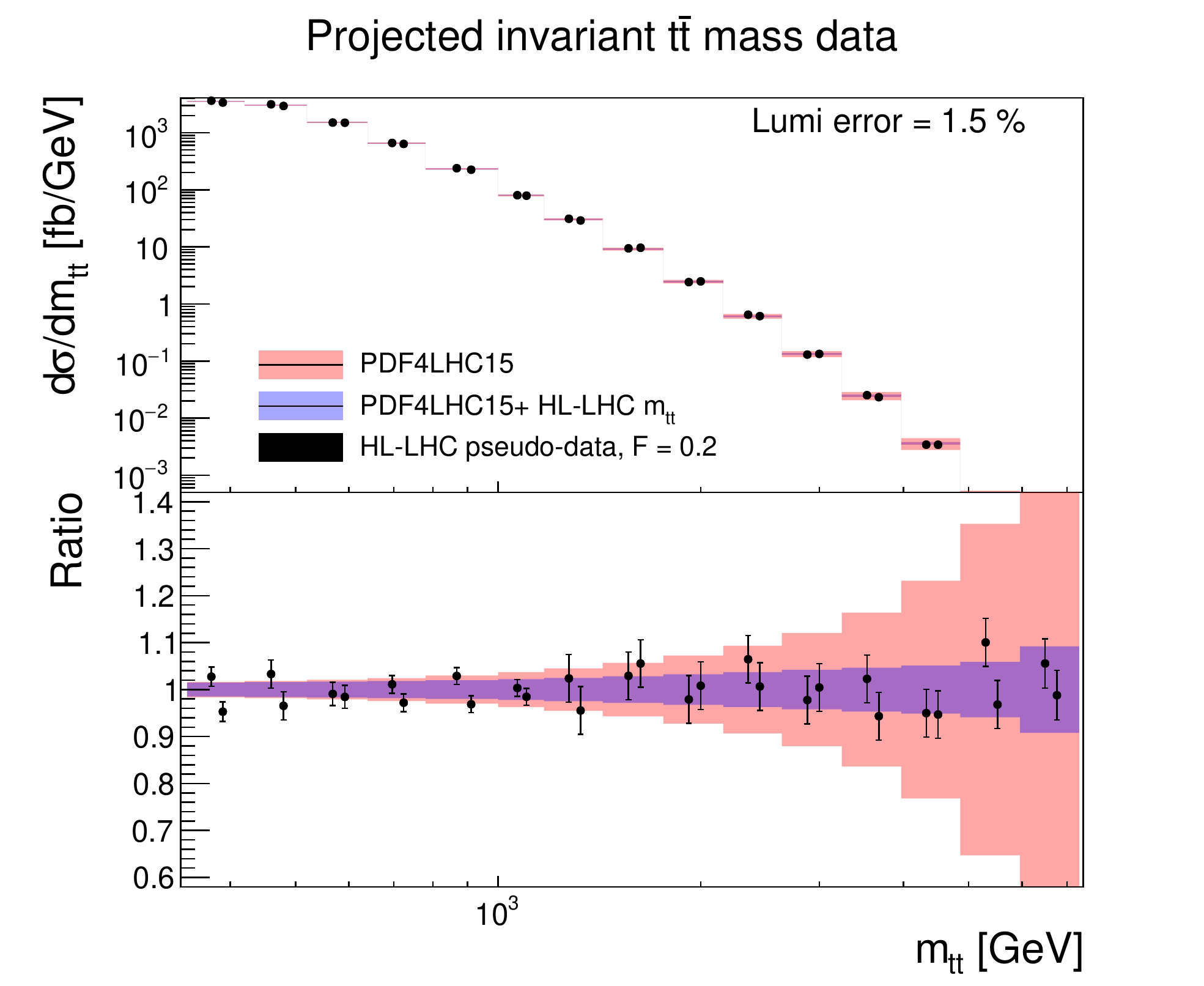}
\end{subfigure}
~
\begin{subfigure}[b]{0.4\textwidth}
   \includegraphics[width=\textwidth]{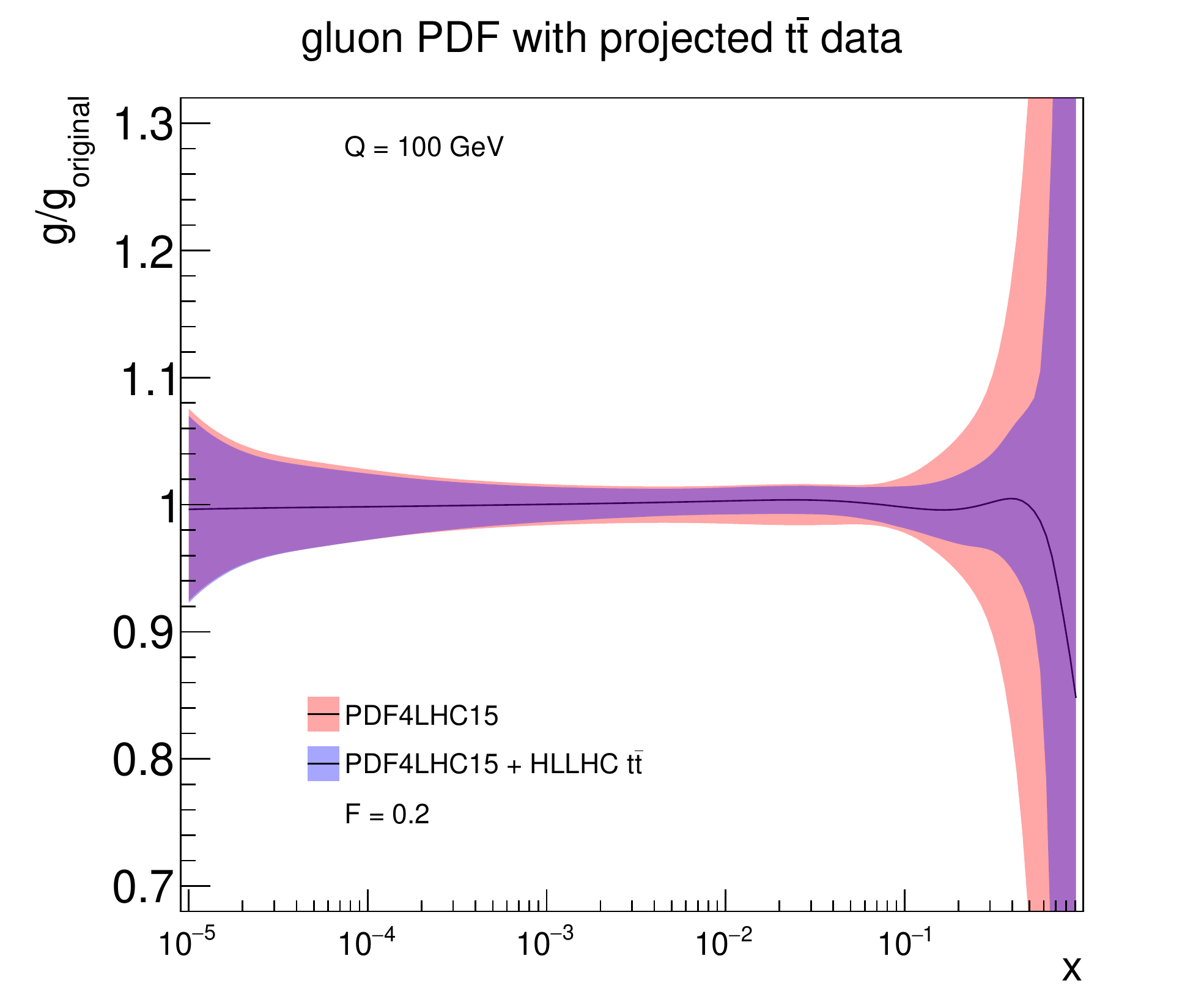}
\end{subfigure}

\caption{Left: Fit to the invariant mass distribution of the $t\bar{t}$ cross section. Right: Effect on the gluon PDF when fitting to multiple differential distributions in the $t\bar{t}$ process.}
\label{individual tt}
\centering
\end{figure}

\noindent Before performing the full fit\footnote{Note: we use the shorthand `fit' for brevity, however it should be understood that a profiling has been preformed, not a full refit.}, we first analyse individual data sets, in order to examine the effects that these have on the PDFs. First we fit to the $t\bar{t}$ pseudo-data sets, as shown in figure \ref{individual tt}. This data set benefits from the extension in the kinematic region, both in transverse momentum and invariant mass. This can be seen as the bins at low invariant mass (previously explored) have negligible improvement but at high invariant mass (unexplored), there is a large error reduction. Overall, this extended reach has allowed the $t\bar{t}$ pseudo-data set to further constrain the high-$x$ gluon, as expected.

\begin{figure}[t]
	\centering
\begin{subfigure}[b]{0.4\textwidth}
  \includegraphics[width=\textwidth]{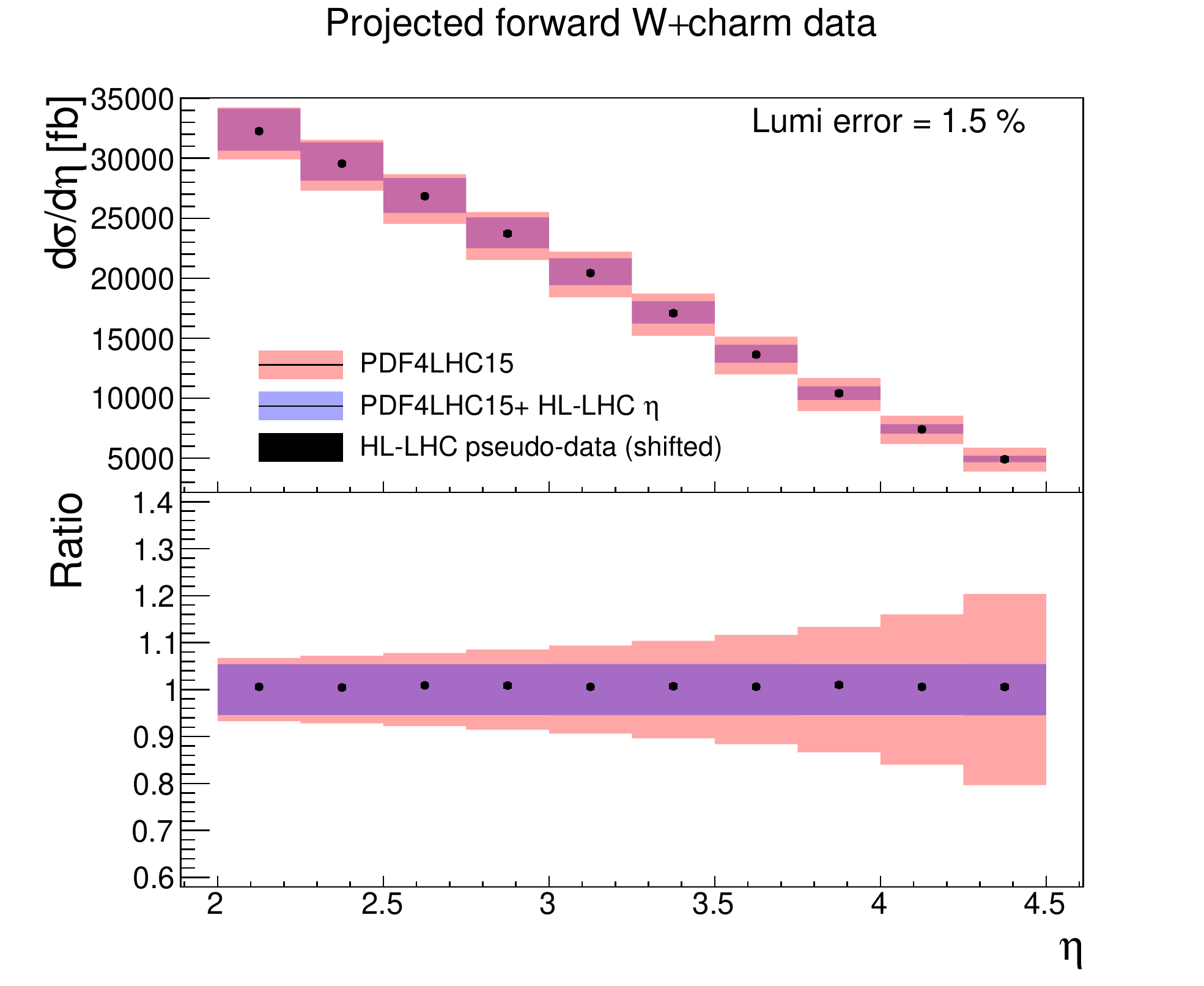}
\end{subfigure}
	~
\begin{subfigure}[b]{0.4\textwidth}
    \includegraphics[width=\textwidth]{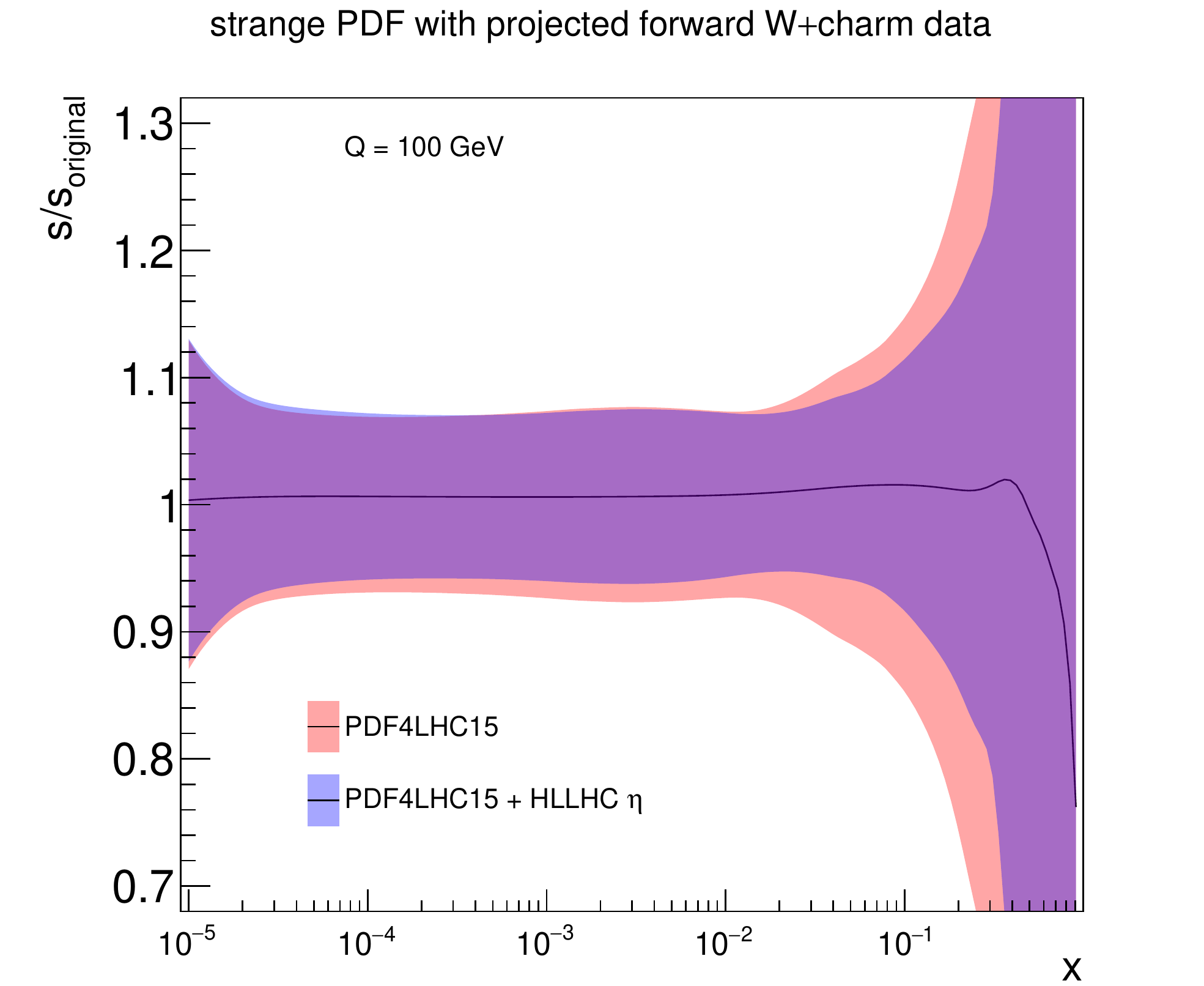}
\end{subfigure}

\caption{Left: Fit to the lepton pseudo-rapidity distribution of the forward $W+$charm cross section. Right: Effect on the strange PDF from the same fit. }
\label{individual}
	\centering
\end{figure}

\noindent Next we examine the forward $W+$charm data set from LHCb. There is currently no data set relating to this process due to the low rate of events and thus a dedicated model with a fully correlated normalisation error is used as suggested by our LHCb colleagues. The results for this are shown in figure \ref{individual}, where we see the statistical error has been reduced such that it is insignificant compared to the normalisation error. Further, this pseudo-data set has a good reduction on the strange quark PDF errors from mid- to high-$x$.

\subsection{Combined Fit}

\begin{figure}[t]
		\centering
	    \begin{subfigure}[b]{0.4\textwidth}
	        \includegraphics[width=\textwidth]{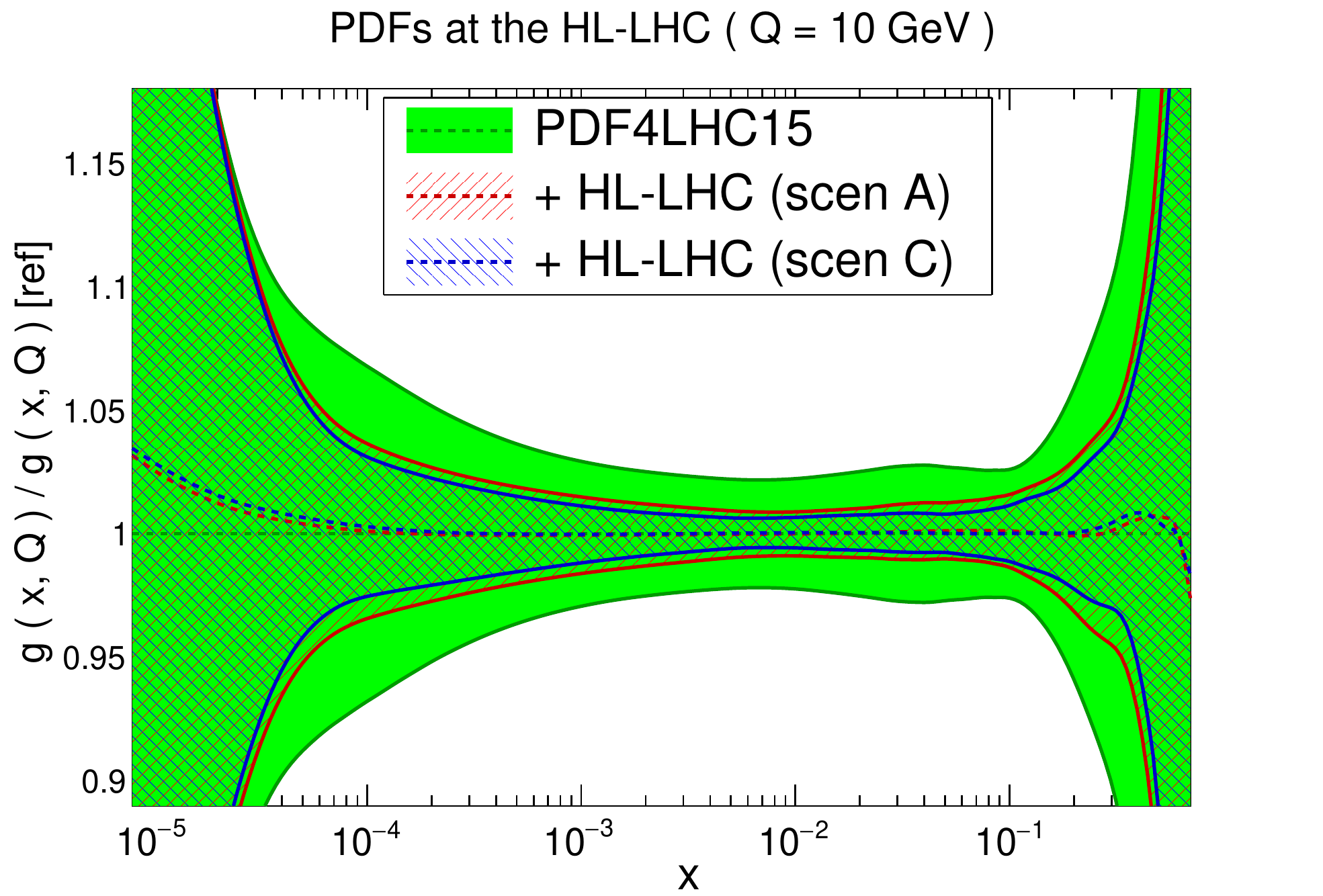}
	    \end{subfigure}
		~
			    \begin{subfigure}[b]{0.4\textwidth}
			        \includegraphics[width=\textwidth]{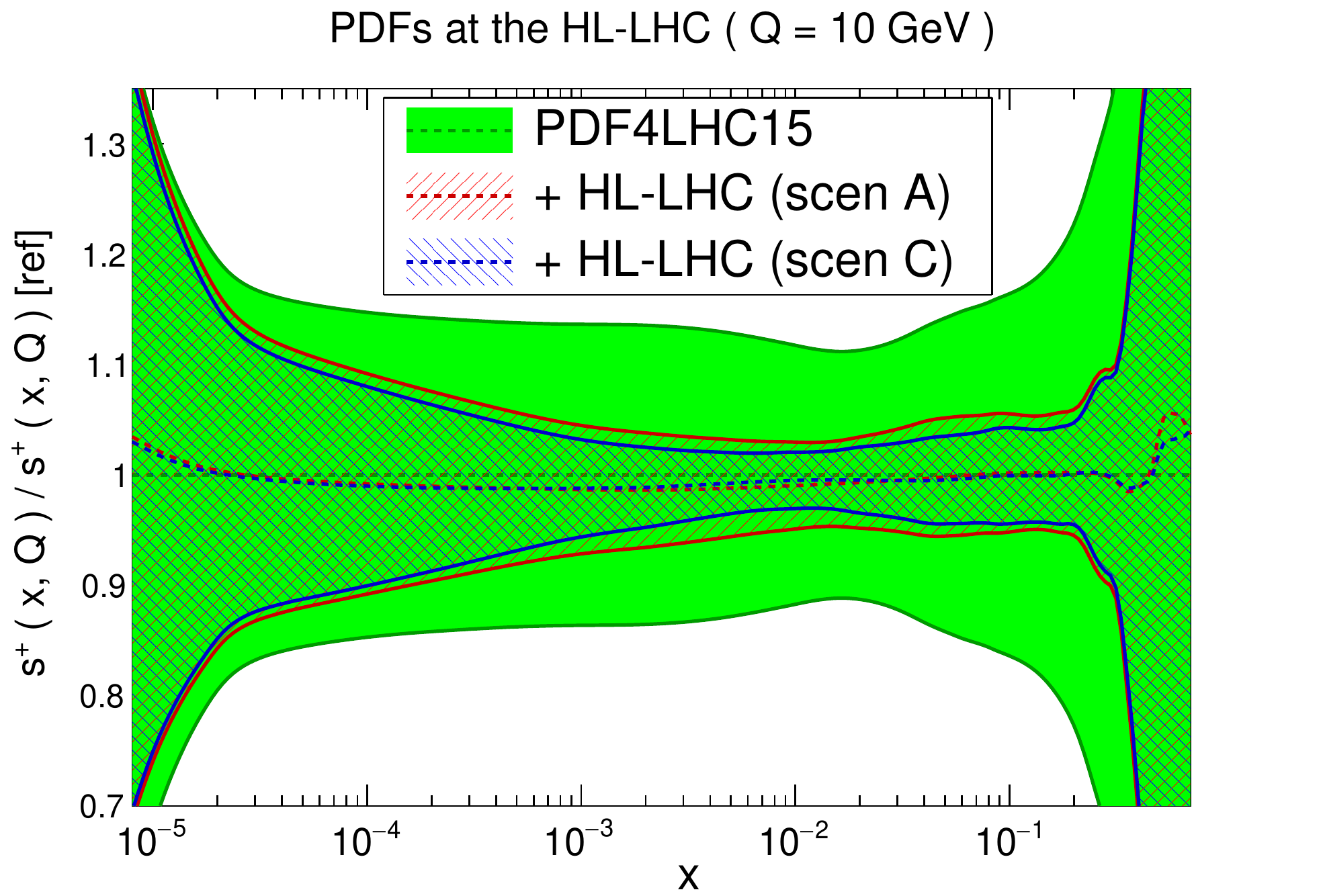}
			    \end{subfigure}

	    \caption{Effects on the gluon (left) and strange (right) PDF when fitting all HL-LHC pseudo-data.}
	    \label{full}
		\centering
	\end{figure}

\noindent We now perform fits to all pseudo-data sets using an optimistic scenario (Scen C: $f_{\text{red}} = 0.1-0.2$) and a conservative one (Scen A: $f_{\text{red}} = 0.5-1.0$). The results for these are shown in figure \ref{full}. We see a good reduction, by a factor of 2-4 for the luminosities, over a wide range of kinematics. In addition, the different scenarios are relatively similar, indicative of the robustness of our methodology. This is to be expected due to our choice of processes likely to benefit most from improved statistics, and thus the exact treatment of the systematic errors has little impact.

\subsection{Impact on SM and BSM Studies}

\noindent A number of SM studies have been performed in the HL/HE-LHC Yellow Report \cite{yellow_report} using our predicted PDFs. For example it has been estimated that the PDF error for the measurement of the $W$ boson mass will be cut in half. Further, one can find a significant improvement in the PDF errors for processes not included in the current fit, for example dijet production.

\begin{figure}[t]
			\centering
		    \begin{subfigure}[b]{0.4\textwidth}
		        \includegraphics[width=\textwidth]{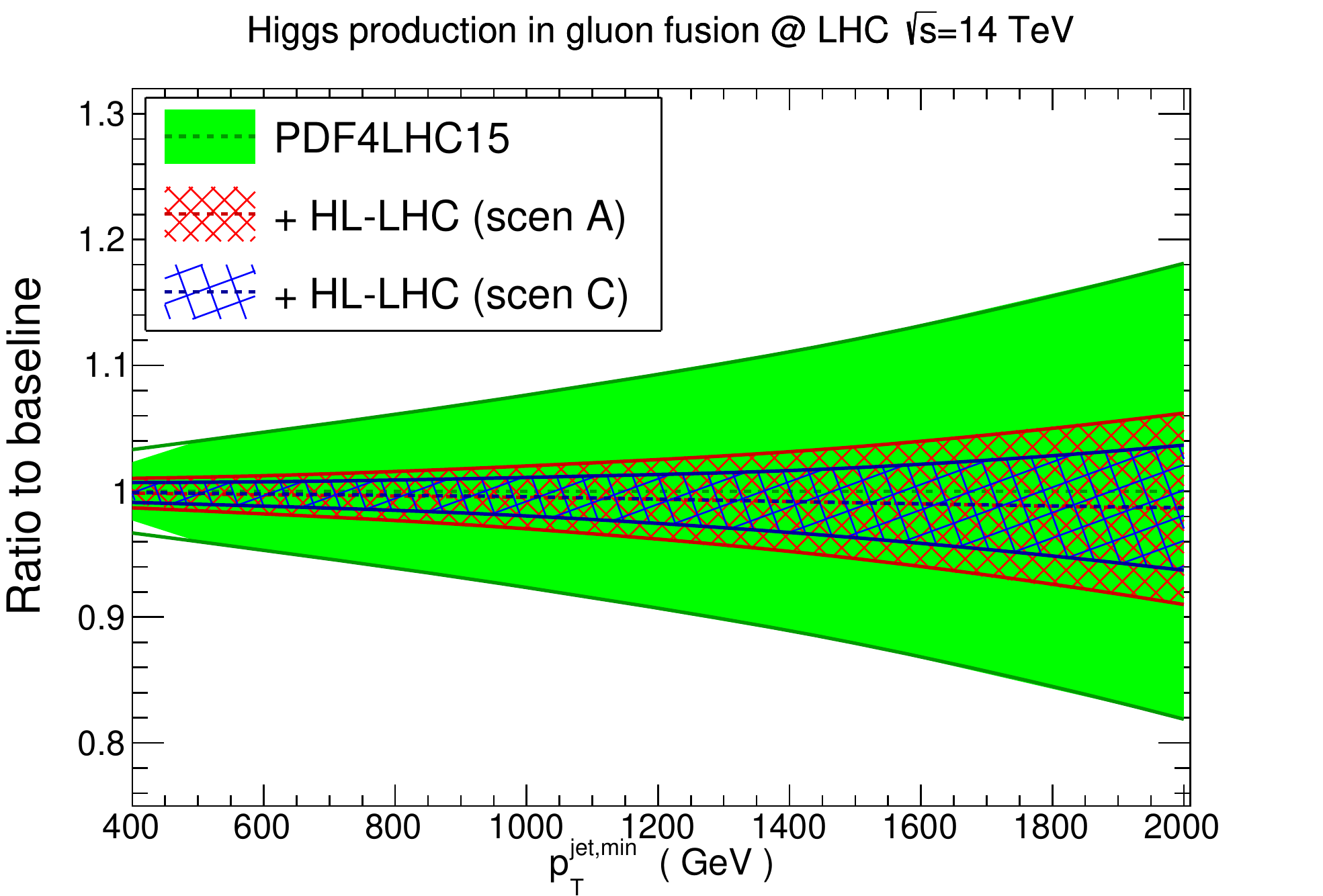}
		    \end{subfigure}
			~
				    \begin{subfigure}[b]{0.4\textwidth}
				        \includegraphics[width=\textwidth]{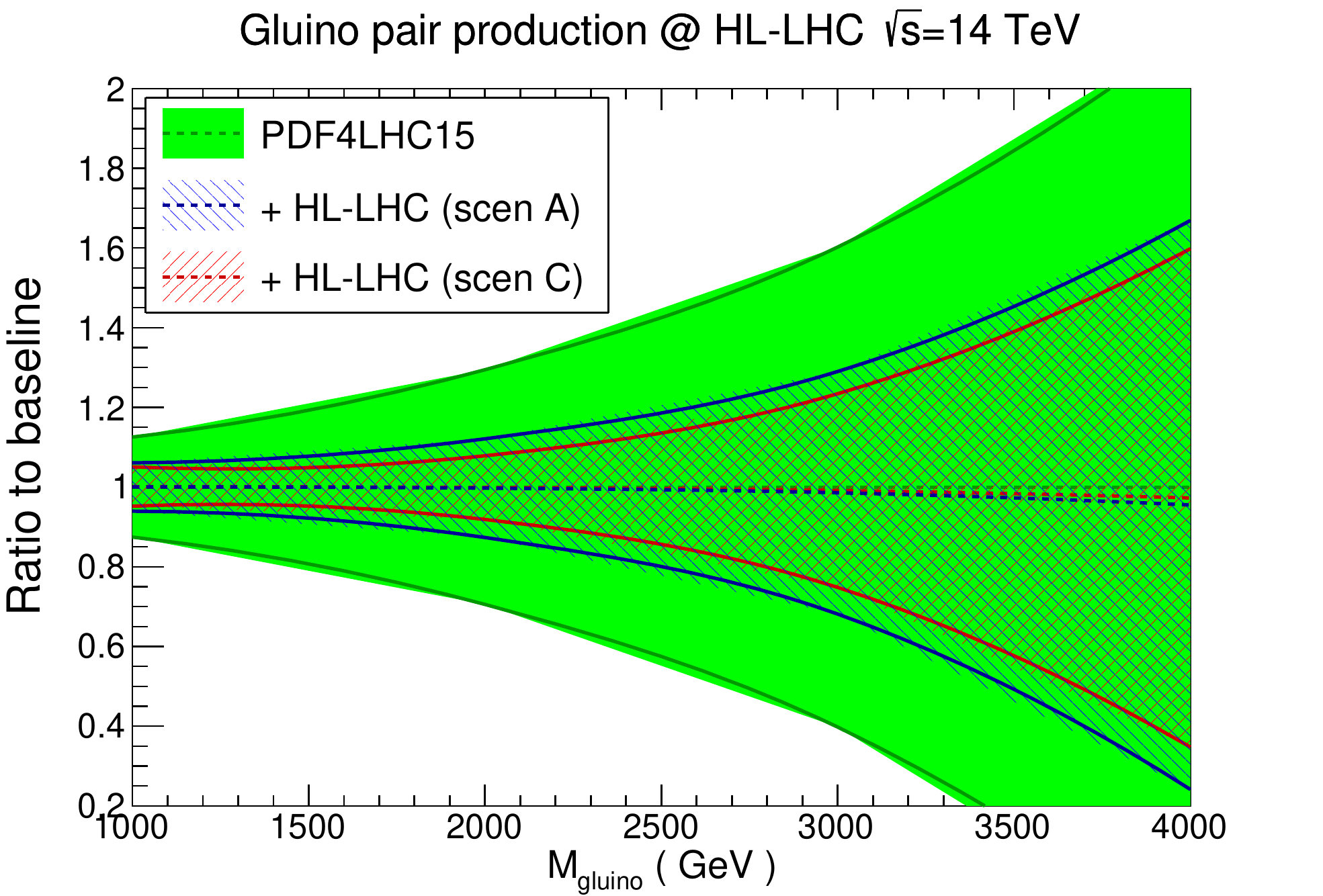}
				    \end{subfigure}

		    \caption{Comparison of PDF errors when calculating the Higgs production via gluon fusion (left) and gluino pair production (right) using the PDF4LHC15 PDF set and the same PDF set fitted to the HL-LHC pseudo-data. }
		    \label{BSM}
			\centering
		\end{figure}

\noindent In addition, we show two standard studies displayed in figure \ref{BSM}. First we look into Higgs production via gluon fusion, where there is a significant reduction, aiding the search for a potential intermediary particle. We also look at gluino pair production where we see good reduction in error over a broad range of kinematics, which will assist us in constraining the parameter space.

\section{LHeC}

\begin{figure}[t]
			\centering
		    \begin{subfigure}[b]{0.4\textwidth}
		        \includegraphics[width=\textwidth]{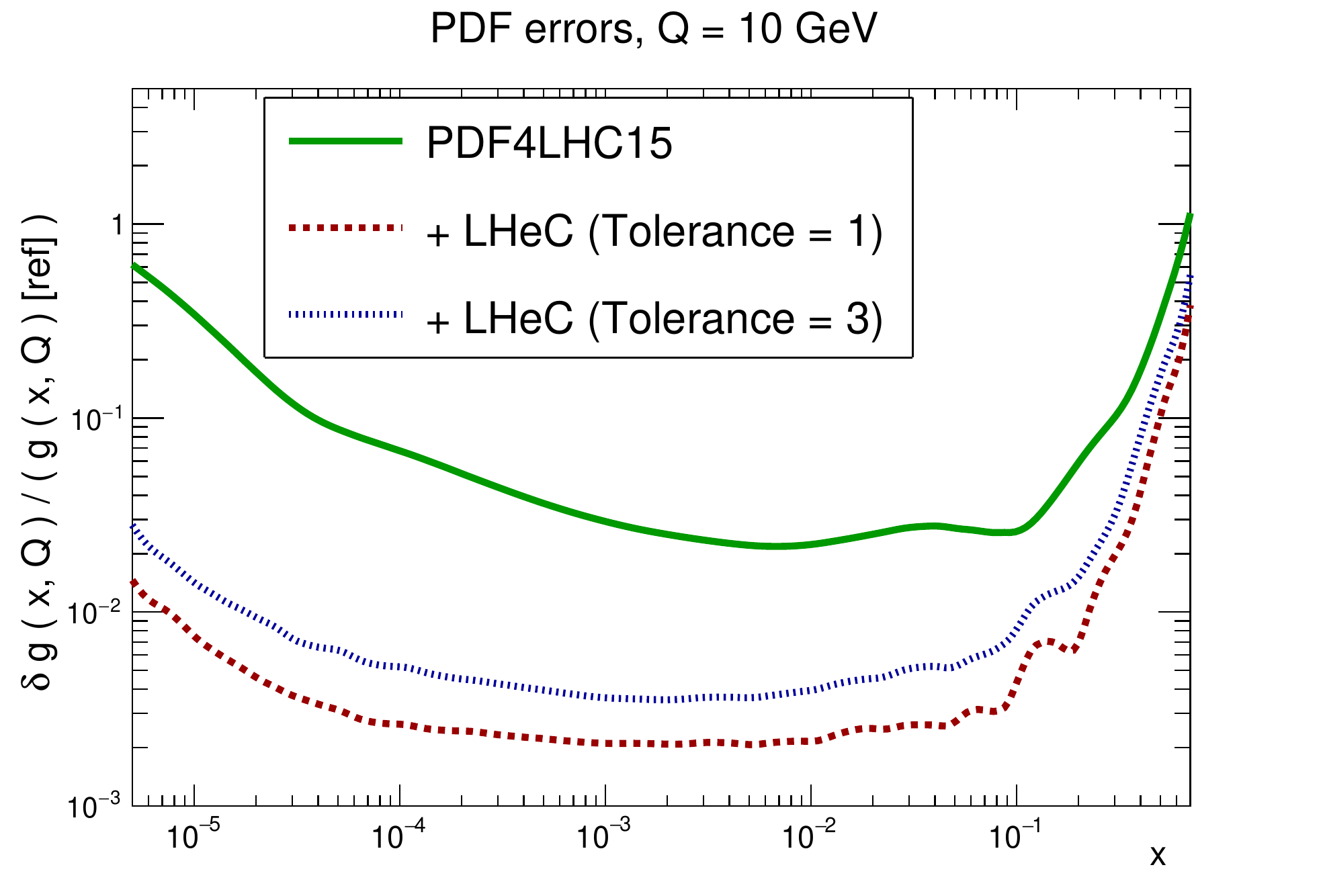}
		    \end{subfigure}
			~
				    \begin{subfigure}[b]{0.4\textwidth}
				        \includegraphics[width=\textwidth]{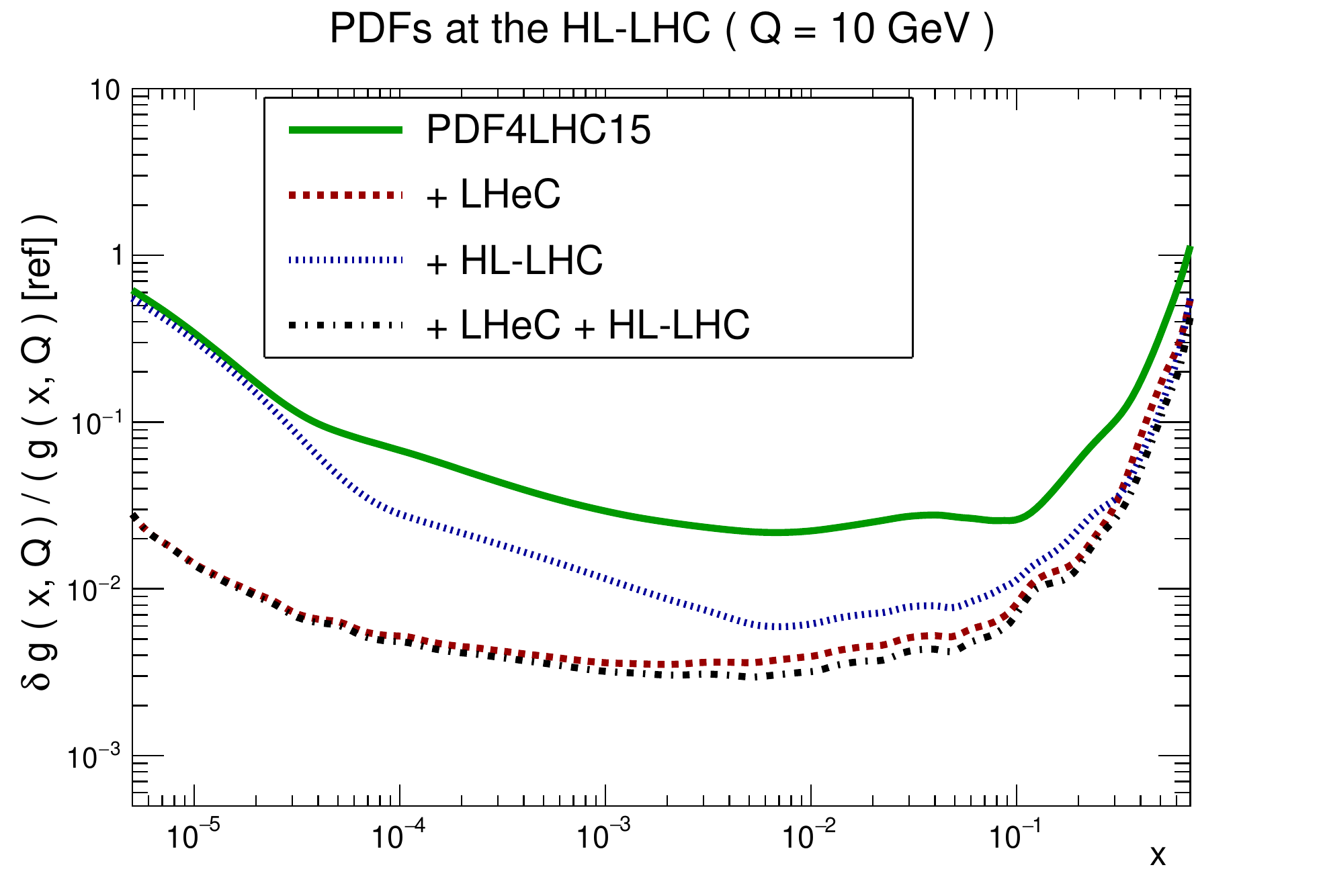}
				    \end{subfigure}

		    \caption{Left: Fits to LHeC pseudo-data using $T$ = 1,3. Right: Combined HL-LHC and LHeC fit. }
		    \label{lhec}
			\centering
		\end{figure}

\noindent We now extend our analysis to the LHeC, and compare the results to the HL-LHC.
For this, we will be using pseudo-data generated in \cite{lhec_data} but ignore the effects of polarisation. The results of this fit are shown in figure \ref{lhec}, where we see that we have significant reduction in errors of a wide range of $x$, in particular at low $x$ as is expected from this data set. One complication that arises is the tolerance, as one may argue that since there is predicted to be less tension between and within each data set at the LHeC, it might not be necessary to set $T=3$ in eq \eqref{eq} and the typical value of unity can be used. However, this would reduce the impact of the baseline PDF and so the exact value of the tolerance to be taken is unclear. Thus, we fit using both $T=1$ and $3$ to compare and note that the impact of doing this is to reduce the overall error by a factor of around 2. 

\noindent Finally, we are able to combine both the HL-LHC and LHeC pseudo-data into one complete fit. We see that both experiments are complementary in their PDF constraining power. However it should be noted that additional data sets could change this picture. For example DIS jets could allow for the LHeC to further constrain the high-$x$ gluon while inclusive $D$ meson production can aid the HL-LHC in impacting the low-$x$ gluon. A more detailed analysis of the impact from the LHeC can be found here \cite{lhec}.

\section{Summary}

\noindent From this analysis we have seen that the High Luminosity phase of the LHC will provide a significant reduction to the PDF errors when compared to current state of the art fits by a factor of 2-4 over a wide range of kinematics. We further see that this leads to improvements of SM and BSM studies by aiding in the reduction of errors on standard quantities, and reducing the available parameter space for new particles. In addition, we showed that the LHeC will have a strong impact on the PDFs, particularly at low-$x$, relevant for DGLAP violation studies. This is also complimentary to the HL-LHC, with different kinematic regions being constrained by either one.
More details included caveats to the methodology used in this study can be found at \cite{hllhc,lhec}.

\bibliographystyle{JHEP}
\bibliography{hllhc_bib}

\providecommand{\href}[2]{#2}\begingroup\raggedright\begin{thebibliography}{1}

\bibitem{yellow_report}
{\scshape HL-LHC, HE-LHC Working Group} collaboration, \emph{{Standard Model
  Physics at the HL-LHC and HE-LHC}},
  \href{https://arxiv.org/abs/1902.04070}{{\ttfamily 1902.04070}}.

\bibitem{hllhc}
R.~Abdul~Khalek, S.~Bailey, J.~Gao, L.~Harland-Lang and J.~Rojo, \emph{{Towards
  Ultimate Parton Distributions at the High-Luminosity LHC}},
  \href{https://doi.org/10.1140/epjc/s10052-018-6448-y}{\emph{Eur. Phys. J.}
  {\bfseries C78} (2018) 962}
  [\href{https://arxiv.org/abs/1810.03639}{{\ttfamily 1810.03639}}].

\bibitem{MCFM}
R.~Boughezal, J.~M. Campbell, R.~K. Ellis, C.~Focke, W.~Giele, X.~Liu et~al.,
  \emph{{Color singlet production at NNLO in MCFM}},
  \href{https://doi.org/10.1140/epjc/s10052-016-4558-y}{\emph{Eur. Phys. J.}
  {\bfseries C77} (2017) 7} [\href{https://arxiv.org/abs/1605.08011}{{\ttfamily
  1605.08011}}].

\bibitem{applgrid}
T.~Carli, D.~Clements, A.~Cooper-Sarkar, C.~Gwenlan, G.~P. Salam, F.~Siegert
  et~al., \emph{A posteriori inclusion of parton density functions in nlo qcd
  final-state calculations at hadron colliders: the applgrid project},
  \href{https://doi.org/10.1140/epjc/s10052-010-1255-0}{\emph{The European
  Physical Journal C} {\bfseries 66} (2010) 503–524}.

\bibitem{NLOJET}
Z.~Nagy, \emph{{Next-to-leading order calculation of three jet observables in
  hadron hadron collision}},
  \href{https://doi.org/10.1103/PhysRevD.68.094002}{\emph{Phys. Rev.}
  {\bfseries D68} (2003) 094002}
  [\href{https://arxiv.org/abs/hep-ph/0307268}{{\ttfamily hep-ph/0307268}}].

\bibitem{pdf4lhc}
J.~Butterworth et~al., \emph{{PDF4LHC recommendations for LHC Run II}},
  \href{https://doi.org/10.1088/0954-3899/43/2/023001}{\emph{J. Phys.}
  {\bfseries G43} (2016) 023001}
  [\href{https://arxiv.org/abs/1510.03865}{{\ttfamily 1510.03865}}].

\bibitem{hessian}
H.~Paukkunen and P.~Zurita, \emph{{PDF reweighting in the Hessian matrix
  approach}}, \href{https://doi.org/10.1007/JHEP12(2014)100}{\emph{JHEP}
  {\bfseries 12} (2014) 100} [\href{https://arxiv.org/abs/1402.6623}{{\ttfamily
  1402.6623}}].

\bibitem{lhec_data}
\url{http://hep.ph.liv.ac.uk/~mklein/lhecdata/} and
  \url{http://hep.ph.liv.ac.uk/~mklein/heavyqdata/}.

\bibitem{lhec}
R.~Abdul~Khalek, S.~Bailey, J.~Gao, L.~Harland-Lang and J.~Rojo, \emph{{Probing
  Proton Structure at the Large Hadron electron Collider}},
  \href{https://arxiv.org/abs/1906.10127}{{\ttfamily 1906.10127}}.

\end{thebibliography}\endgroup

\end{document}